\documentclass[reprint,aps,prl,amsmath,nofootinbib,sort&compress]{revtex4-1}

\usepackage{xcite}
\usepackage{graphicx}
\usepackage{dcolumn}
\usepackage{bm}

\usepackage{amsmath,amssymb,amstext}
\usepackage{subfigure}

\usepackage{float}
\usepackage{color, colortbl}
\usepackage[table,xcdraw]{xcolor}
\definecolor{LightCyan}{rgb}{0.88,1,1}

\usepackage{caption}
\usepackage[utf8]{inputenc}
\usepackage{lmodern}

\usepackage{mathtools,nccmath}%

\usepackage{environ}

\usepackage{hyperref}

\usepackage{graphicx} 

\usepackage{xcolor}

\graphicspath{{./}{figs/}}

\begin{document}


\title{Relativistic Asymmetric Reconnection}

\author{Rostom Mbarek}
 \email{rmbarek@uchicago.edu}
\affiliation{%
Department of Astronomy \& Astrophysics, University of Chicago, Chicago, IL 60637, USA
}%
\affiliation{Kavli Institute for Cosmological Physics, The University of Chicago, Chicago, IL 60637, USA}
\affiliation{Enrico Fermi Institute, The University of Chicago, Chicago, IL 60637, USA}

\author{Colby Haggerty}%
\affiliation{%
Institute for Astronomy, University of Hawai`i, Honolulu, HI 96822, USA
}%
\affiliation{%
Department of Astronomy \& Astrophysics, University of Chicago, Chicago, IL 60637, USA
}%

\author{Lorenzo Sironi}%
\affiliation{%
Department of Astronomy \& Columbia Astrophysics Laboratory, Columbia University, New York, NY 10027, USA
}

\author{Michael Shay}%
\affiliation{%
Bartol Research Institute, Department of Physics \& Astronomy, University of Delaware, Newark, DE, 19716, USA
}%

\author{Damiano Caprioli}
\affiliation{%
Department of Astronomy \& Astrophysics, University of Chicago, Chicago, IL 60637, USA
}%
\affiliation{Enrico Fermi Institute, The University of Chicago, Chicago, IL 60637, USA}

\date{\today}

\begin{abstract} 
We derive basic scaling equations for relativistic magnetic reconnection in the general case of asymmetric inflow conditions and obtain predictions for the outflow Lorentz factor and the reconnection rate. Kinetic Particle-in-Cell simulations show that the outflow speeds as well as the nonthermal spectral index are constrained by the inflowing plasma with the weaker magnetic energy per particle, in agreement with the scaling predictions. 
These results are significant for understanding non-thermal emission from reconnection in magnetically-dominated, astrophysical systems, many of which may be asymmetric in nature.
The results provide a quantitative approach for including asymmetry on reconnection in the relativistic regime.
\end{abstract}

\maketitle

\paragraph{\underline{Introduction}}

Magnetic reconnection is a fundamental plasma process through which energy stored in magnetic fields is converted into thermal and nonthermal particle energy. 
This process 
is a candidate for explaining impulsive nonthermal emission from magnetically dominated astrophysical objects such as gamma-ray bursts (GRBs)
\citep[e.g.][]{zhang-yan11,mcKinney-uzdensky11},
pulsar winds \citep[e.g.][]{lyubarsky-kirk01,arons12}, and jets from active galactic nuclei \citep[e.g.][]{giannios10,sironi+21}.
As the magnetic energy density becomes larger than the rest mass energy density of the plasma, reconnection generates power law distributions of energetic ($>{\rm MeV}$) leptons \citep{guo+14,guo+15,sironi+15,werner+16}. 

There have been significant efforts to understand the basics of relativistic reconnection including: (i) its dynamics and scaling \cite{blackman+f94,lyubarsky05,lyutikov03,takahashi+11}, (ii) the rate at which magnetic energy is dissipated, i.e., the reconnection rate \cite{zenitani+ho07,hesse+11,liu+15}, and (iii) nonthermal particle acceleration \cite{zenitani+ho01,zenitani+ho05,sironi+sp11,sironi+sp14,liu+15,zenitani+he08,jaroschek+04,kagan+13,kagan+15,bessho+bh12,cerruti+12,cerruti+13,cerruti+14a,cerruti+14b,rowan+17,ball+18}.
These works have shown that relativistic reconnection efficiently accelerates nonthermal particles.
However, such studies have assumed that the magnetic field strength, temperature and density of the reconnecting plasma regions are equal (symmetric case). 
Nevertheless, systems in which the inflowing parameters differ, i.e., asymmetric reconnection, is not an odd or rare event, with in-situ observations of asymmetric reconnection frequently reported in the heliosphere \citep{oieroset+04,paschmann+13,phan+13,burch+16,mistry+17}.
Because of the prevalence of asymmetric reconnection in near-Earth systems, asymmetric reconnection has been thoroughly detailed \cite{swisdak+03,cassak+sh08,malakit+10,shay+16}.
Albeit extensive, this body of work is limited to non-relativistic systems, and presently, the description for asymmetric reconnection has not been extended to relativistic plasma, i.e., systems where the magnetic energy density is larger than the rest mass energy density of the plasma.
The ubiquity of asymmetric reconnection in the heliosphere suggests that asymmetric reconnection can be as important to astrophysical environments.
Likely examples include the boundary layer between the jet and accretion flow in active galactic nuclei (AGNs) or shear flow boundaries warped by Kelvin-Helmholtz instabilities in AGN jets \cite[e.g.][]{lyubarsky-kirk01,ripperda+20,sironi+21}.

With this in mind, we present the first step in understanding asymmetric reconnection in the relativistic regime.
We derive basic scaling predictions for relativistic asymmetric reconnection 
which reproduce both the symmetric relativistic and non-relativistic asymmetric limits.
We show that for asymmetric inflow density and magnetic fields, the reconnection rate and the outflow speed are set by the inflow with the weaker ratio of magnetic energy density to rest mass energy density, i.e., the magnetization $\sigma$.
The predictions are tested with a survey of two-dimensional (2D) particle-in-cell (PIC) simulations, and show good agreement.
Finally, we examine the effects of inflowing asymmetry on nonthermal particle acceleration, and show that the efficiency of particle acceleration is again controlled by the weaker-magnetization region.

\paragraph{\underline{Scaling of Asymmetric Reconnection}}

If reconnection is occurring at a steady state, the characteristics of the outflowing plasma can be related to the inflowing properties by enforcing the conservation of mass and energy around a given X-line.\footnote{Note that the conservation of momentum flux only provides a pressure balance constraint, and does not connect the inflow properties to to the outflow one.}
To this end, we consider an asymmetric diffusion region,
with two distinct plasmas flowing in from below and above the current sheet; each with its own density and magnetic field strength flowing into a side with length $L$ and with bulk velocities $v_1$ and $v_2$ (the 1 \& 2 subscripts denote the distinct regions below and above the current sheet respectively).
The two populations mix as the magnetic energy is dissipated and the plasma is accelerated out of either side of the diffusion region with width $\delta$. For simplicity, we assume that the temperature of the inflowing plasma is non-relativistic and the inflowing magnetic fields are anti-parallel, i.e., the guide field is negligible.
 
The field on both sides of the current sheet is characterized by the magnetization, defined in this paper as $\sigma \equiv B^{\prime 2}/4 \pi \rho^\prime c^2$ where
$B^\prime$ is the magnetic field strength, $\rho^\prime$ is the mass density, and $c$ is the speed of light.
Note that throughout the paper, unless otherwise stated, un-primed variables are measured in the X-line (or simulation) frame and primed variables are measured in the rest frame, i.e. the frame co-moving with the fluid velocity $v$ with Lorentz factor $\gamma = 1/\sqrt{1 - (v/c)^2}$. 

The equation governing the conservation of mass in the X-line frame is:
\begin{equation}\label{m_cons}
    L (\gamma_1 \rho^\prime_1 v_1 + \gamma_2 \rho^\prime_2 v_2) = 2 \gamma_{\rm out} \rho^\prime_{\rm out}  v_{\rm out} \delta
\end{equation}
The conservation of energy density flux \citep{Lichnerowicz+67,lyutikov+03} yields:
\begin{equation}\label{e_cons}
\begin{split}
L\left [\left (w^\prime_1 + \frac{B^{\prime 2}_1}{4\pi}\right )\gamma_1^2 v_1 + 
        \left (w^\prime_2 + \frac{B^{\prime 2}_2}{4\pi}\right )\gamma_2^2 v_2 \right]  \\
    =2 \delta\left (w^\prime_{\rm out} + \frac{B^{\prime 2}_{\rm out}}{4\pi}\right )\gamma_{\rm out}^2 v_{\rm out},
\end{split}
\end{equation}
where $w^\prime = \rho^\prime c^2 + \Gamma P_0^\prime /(\Gamma - 1)$ is the enthalpy density, $\Gamma$ is the adiabatic index of the plasma, and $P^\prime$ is the pressure.
We take the reconnection region to be in steady state and apply Stokes' theorem to Faraday's law to obtain $\vec{\nabla} \times \vec{E} = 0$; as a result, the inductive electric field driving reconnection $E_z$ is uniform. The inflowing velocities are expressed as $v_i/c = |\vec{E} \times \vec{B}_i|/B^{2}_i \approx E/B_i$, implying that $\gamma_1 v_1B^\prime_1 = \gamma_2v_2B^\prime_2$, since $E_z$ is the same on both sides of the current sheet.

Using these considerations, assuming that $B^{\prime}_{\rm out}$ is negligible compared to $B^{\prime}_{1}$ and $B^{\prime}_{2}$, and dividing Eq. \ref{e_cons} by Eq. \ref{m_cons}, we obtain,
\begin{equation}
 \left ( 1 + \frac{\Gamma P^\prime_{\rm out}}{(\Gamma - 1) \rho^\prime_{\rm out} c^2}\right) \gamma_{\rm out} =
 \frac{\gamma_1(1 + \sigma_1) + \xi \bf{\gamma_2} (1 + \sigma_2)}
 {1 + \xi}\label{eq:PRhoGamma}
\end{equation}
where $\xi \equiv \frac{B^\prime_1 \rho^\prime_2}{B^\prime_2 \rho^\prime_1} = \sqrt{\frac{\sigma_1\rho^\prime_2}{\sigma_2\rho^\prime_1}}$ and the inflowing initial temperatures are assumed to be non-relativistic, $w^\prime_{1,2} \approx \rho^\prime_{1,2}c^2$.
This equation provides a prediction for the Lorentz factor of the outflowing plasma $\gamma_{\rm out}$, provided we know the outflowing pressure and density.
Previous works on the scaling of reconnection have made diverging assumptions about the role of the thermal pressure in the exhaust:
some works have assumed the pressure is negligible compared to the outflowing bulk kinetic energy density \citep[e.g.][]{cassak+sh08,lyutikov03}, while others take the pressure to be relativistically hot and comparable to the outflowing energy density \citep[e.g.][]{lyubarsky05,zenitani+he08}.
For completeness, in this work we present both cases and show that simulations agree with the hot exhaust scenario.

In the limit of negligible pressure in the exhaust, $w^\prime_{\rm out} \approx \rho^\prime_{\rm out} c^2$, i.e. the magnetic energy is fully utilized to accelerate the plasma, Eq.~\ref{eq:PRhoGamma} becomes
\begin{equation}\label{asym_rrate_ICP}
    \gamma_{\rm out} = \frac{\gamma_1(1 + \sigma_1)  +  \gamma_2\xi(1+ \sigma_2)}{1 + \xi},
\end{equation}
This equation can be verified by considering the limiting behaviour. 
For the symmetric relativistic case, as $\sigma_1 = \sigma_2 = \sigma_{\rm in}$ and $\xi \rightarrow 1$, Eq.~\ref{asym_rrate_ICP} becomes  $\gamma_{\rm out} \approx \gamma_{\rm in}(1 + \sigma)$, identical to the super-Alfv\'enic prediction in \citet{lyutikov+03}; in the non-relativistic limit, we obtain the familiar $v_{\rm out} \sim v_A = B_{\rm in}/\sqrt{4\pi\rho_{\rm in}}$.
Next, we consider the non-relativistic asymmetric limit. Expanding Eq.~\ref{asym_rrate_ICP} in the $v_{\rm out} \ll c$ limit, we find $v_{\rm out}^2 \approx 2 c^2(\sigma_1 + \xi \sigma_2)/(1 + \xi)$, which can be rewritten in terms of the magnetic fields and densities: $v_{\rm out}^2 \approx B_1B_2/(2\pi \rho_{\rm tot})$, where $\rho_{\rm tot} = (\rho_1B_2 + \rho_2B_1)/(B_1 + B_2)$, in agreement with the well-established predictions for non-relativistic asymmetric outflows \cite{cs07,Birn+10}.

In the alternative case, in which the outflowing pressure is much larger than the rest mass energy density and cannot be neglected in the scaling derivation, an expression is needed for the exhaust pressure. 
In non-relativistic magnetic reconnection, the energy per particle in the bulk flow is found to be comparable with the exhaust temperature in simulations, observations and experiments \cite{shay+14,haggerty+17,haggerty+18,phan+13,phan+14,yamada+15}. 
Motivated by this and by results from the simulations preformed for this work, the exhaust pressure is taken to be, 
$P^\prime_{\rm out} \sim \rho^\prime_{\rm out}\gamma_{\rm out}v_{\rm out}^2 \sim \rho^\prime_{\rm out}\gamma_{\rm out}c^2$ ,
and in this limit, the left hand side of Eq.~\ref{eq:PRhoGamma} is equal to 
$4\gamma_{\rm out}^2$, where we have used the relativistic adiabatic index of $\Gamma = 4/3$. Applying this to Eq.~\ref{eq:PRhoGamma} yields
\begin{equation}\label{asym_rrate}
    \gamma_{\rm out} \approx \sqrt{\frac{\gamma_1(1 + \sigma_1)  +  \gamma_2\xi(1+ \sigma_2)}{4(1 + \xi)}}.
\end{equation}
Taking this equation in the symmetric case, we recover the scaling relation of $\gamma_{\rm out} \propto \sqrt{\sigma}$, consistent with the Alfv\'enic outflow predictions of \citet{lyubarsky05} and \citet{liu+15}.

Next we consider the limiting behaviour of Eq.~\ref{asym_rrate} for  $\sigma_2 \gg \sigma_1 \gg T/mc^2$.
In this limit, balancing magnetic and thermal pressure requires that $\rho^\prime_1/\rho^\prime_2 \sim \sigma_2/\sigma_1$, $B^\prime_2 \sim B^\prime_1$ (hence $\gamma_1 \sim \gamma_2$), and $\xi \sim \sigma_1/\sigma_2  \ll 1$, the prediction for the outflowing Lorentz factor (Eq. \ref{asym_rrate}) becomes:
\begin{equation}\label{quench}
    \gamma_{\rm out} \sim \frac{\sqrt{\gamma_1(1 + 2\sigma_1)}}{2}
\end{equation}
Thus, for $\sigma_2 \gg \sigma_1$, the outflow speed is set by the weaker-$\sigma$ side, or equivalently the larger density side. This prediction should hold for both relativistic and non-relativistic values of $\sigma_1$.

Finally, we can use the outflow prediction to estimate the reconnection rate $R \equiv v_{1}/v_{\rm out}$, which is $ \approx v_2/v_{\rm out}$, for systems where the upstream thermal energy is much smaller than the magnetic energy per particle. 
In such systems, $v_1 \sim v_2$ since $B^\prime_2 \sim B^\prime_1$ and $\xi \sim \sigma_1/\sigma_2$, and hence, we can rewrite Eq.~\ref{m_cons} in terms of the reconnection rate. 
In the limit where the inflow velocity is not ultra relativistic ($\sigma < 100$ and $\gamma_{\rm in} \gtrsim 1$), we find an approximation for the asymmetric reconnection rate by taking the exhaust density as $\rho_{\rm out} \sim (\rho_1 + \rho_2)/2$ following \citet{cs07}:
\begin{equation}\label{eq:rec_rate}
    R \approx \frac{\delta}{2 L}\sqrt{1 + \frac{2\sigma_1\sigma_2}{\sigma_1 + \sigma_2}}
\end{equation}
This prediction is again only valid for mildly relativistic cases ($\gamma_{\rm in} \gtrsim 1$); nevertheless, considering that the weaker-$\sigma$ side sets the reconnection rate, such a   scaling is pertinent for asymmetric environments with min($\sigma_1, \sigma_2 )\lesssim$ 100, and consequently for symmetric environments with $\sigma \lesssim$ 100.

\paragraph{\underline{Simulation Setup}}
To test the scaling predictions outlined above, we perform 2.5D (2D real space, 3D velocity space) particle-in-cell (PIC) simulations of electron-positron (pair-plasma) reconnection with TRISTAN-MP \citep{bunemann93,spitkovsky05}.
The plasma is comprised of electron-positron pairs, typical of high-energy astrophysical environments such as pulsar winds and AGN jets \citep[e.g.][]{sturrock71,wardle+98}. 
Both electrons and positrons are initialized with a constant temperature, $\Delta \gamma \equiv k_BT^\prime/mc^2 = 0.25 $. 
Throughout the simulations, no notable differences between electron and positron properties are observed.

Initializing a single stable current sheet in the simulations requires  balancing pressure across the current sheet; and thus the density and magnetic field on the two sides are uniquely determined for a fixed temperature and a particular combination of $\sigma_1$ and $\sigma_2$\footnote{The ratio of the magnetic fields across the exhaust is given by 
 $B^\prime_2/B^\prime_1 = \sqrt{\frac{1 + 4\Delta \gamma/\sigma_1}{1 + 4\Delta \gamma/\sigma_2}}$.}. 
Note that pressure balance stipulates that the magnetic field is roughly similar across the sheet if the initial temperature is relatively low as discussed above. 
For additional details on the simulations, the setup and the normalization see the Supplemental Material.
 
\begin{figure}
\centering
\includegraphics[width=0.48\textwidth, clip,trim= -10 0 0 0]{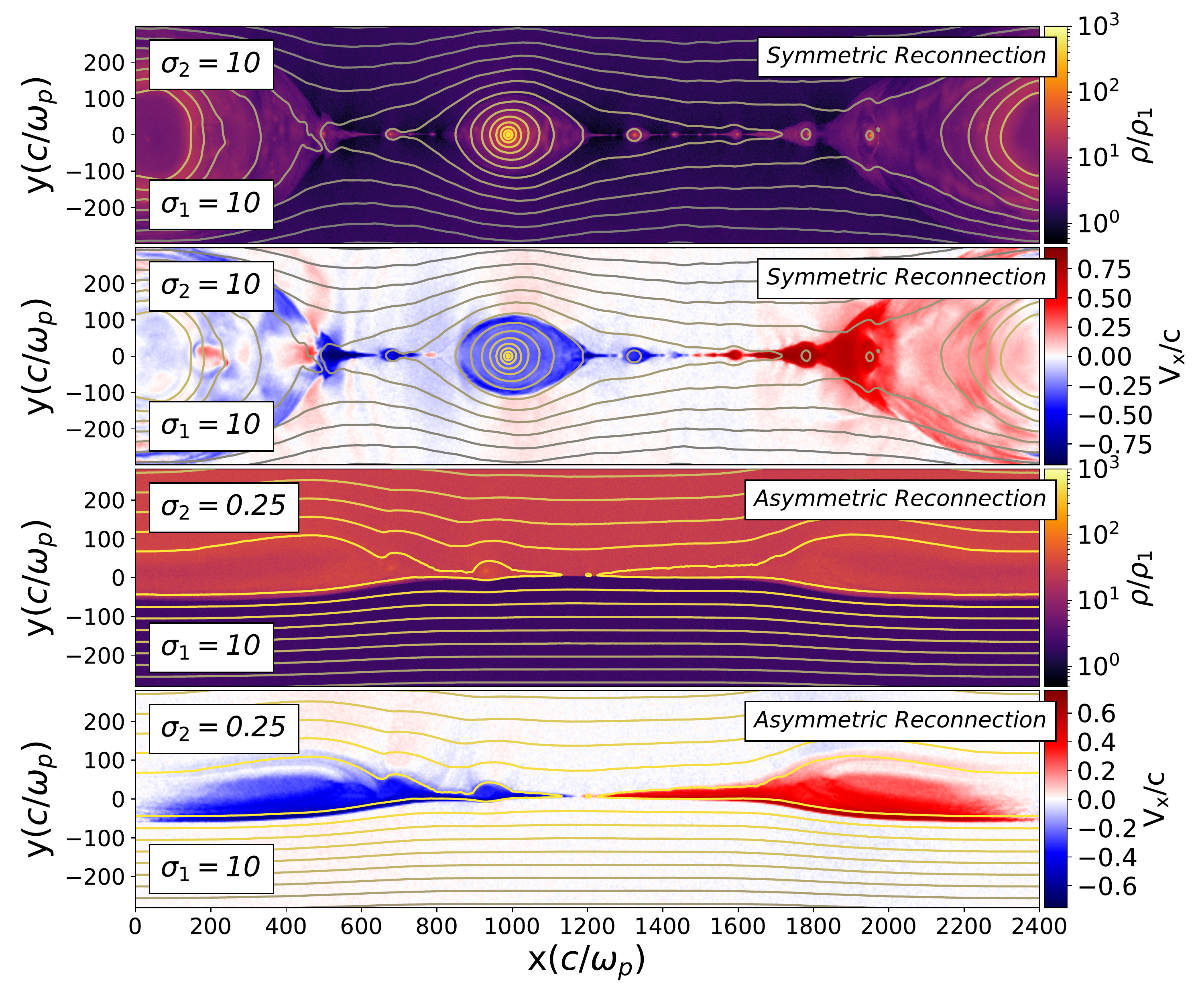}
\caption{
Reconnection layer at $t \omega_p \sim 400$ for both symmetric and asymmetric reconnection. 
The magnetization $\sigma$ is specified in the plot on both sides of the current sheet. Contour maps of the magnetic scalar potential are also shown for reference. Top Panels: mass density map normalized to the mass density in the lower region $\rho_1$. 
Bottom Panels: Outflow speed $v_x$ in units of c.}
\label{sim}
\end{figure}
\paragraph{\underline{Simulation Results}} We study relativistic asymmetric reconnection by performing a survey of simulations where the upper 
magnetization
scans between $\sigma_2 = 10^{-1}$ and $10^3$; the lower is fixed at $\sigma_1 = 10$.
Fig.~\ref{sim} compares 
the mass density ($\rho$) and outflow velocities ($v_x$) for the $\sigma_2 = 10$ case (i.e. symmetric) and the $\sigma_2 = 0.25$ case (i.e. asymmetric), at $t \omega_p\approx 400$ for positrons. Note, that for the latter asymmetric case the assumption $\sigma_2 \gg T/mc^2$ is not satisfied, however we include this limit to show the effect two reconnecting regions with $\sigma$'s that differ by orders of magnitude.
The most notable difference between the symmetric and asymmetric cases is the shape of the exhaust that morphs from round islands to a more elongated shape, preferentially protruding into the inflow region with the weaker magnetic field strength because it has a faster inflow velocity. 
The island 
bulges into this side to replace the higher volume of plasma that has reconnected \citep{krauss+99,cs07}.
Additionally, more plasmoids are present in the symmetric simulation compared to the asymmetric case, which may be due to a combination of the apparent broader asymmetric exhaust and the documented reduction of plasmoids for weaker $\sigma$ in symmetric reconnection \cite{sironi+16}. 
Finally, the area of the islands suggest that less magnetic flux has been reconnected in the asymmetric simulation, consistent with the scaling predictions associated with Eq.~\ref{asym_rrate}.
For the $\sigma_2 \gg \sigma_1$ case (not shown), the shape of the exhaust looks similar to symmetric reconnection, with the island expanding both inflow regions equally.
\begin{figure}
\includegraphics[width=0.45\textwidth, clip,trim=10 10 10 0]{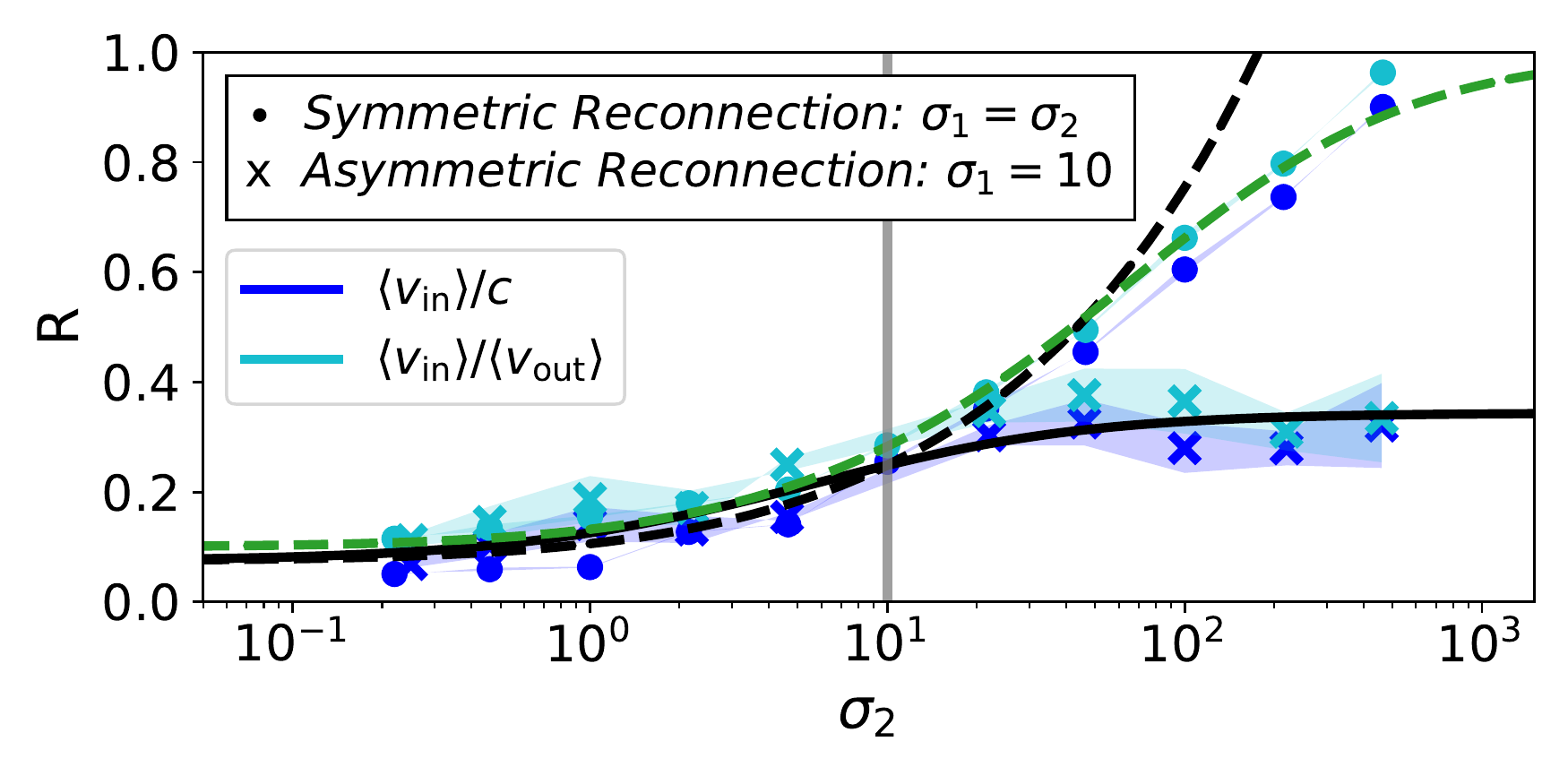}
\caption{ Reconnection rate for different magnetization parameters in simulations of symmetric and asymmetric cases.
The green dashed curve shows the expected symmetric reconnection rate based on \citet{liu+15}'s prediction including the plasma thermal pressure.
The teal and blue dots (crosses) show the averaged measured values of $\langle v_{\rm in} \rangle $ normalized to $\langle v_{\rm  out} \rangle$ and $c$ respectively for symmetric (asymmetric) reconnection.
The averages are determined over a $5d_e$ window just upstream of the X-line.
The black dashed (solid) line shows the prediction for the for symmetric (asymmetric) reconnection rate determined by Eq.~\ref{eq:rec_rate}; such a prediction is only valid for environments with min($\sigma_1,\sigma_2) \lesssim 100$ as explained in the text.
For the asymmetric cases, $\sigma_1 = 10$ is fixed and the reconnection rate is controlled by the weaker $\sigma$.}
\label{combined}
\end{figure}
 
To further verify the scaling predictions, the reconnection rates in the survey of symmetric and asymmetric simulations are measured and shown in Fig.~\ref{combined} as a function of $\sigma_2$.
The teal dots and crosses show $v_{\rm in}/v_{\rm out}$, where $v_{\rm out}$ is a spatially averaged absolute value in the center of the exhaust ($y \sim 0$) and $v_{\rm in}$ is the absolute value of the average of both sides within $5 c/\omega_{p}$ away from the X-line in the inflow region. 
The blue dots and crosses show a similar value, but with the outflow velocity assumed to be $c$ (appropriate for $\sigma_1,\sigma_2 \gg 1$); 
The dashed green line shows the prediction for the reconnection rate from Eq.~5 of \citet{liu+15}
\footnote{For Eq.~3 from \citet{liu+15}, we use $r_{n^\prime}\delta/L=0.1$ and include the thermal contribution to enthalpy with $k_BT'=mc^2/4$.}.
The black dashed and solid lines show the expected reconnection rate based on Eq.~\ref{eq:rec_rate} for $\delta/L = 0.15$ (the best fit for our simulation results), which is comparable with \citet{liu+15}'s favored aspect ratio.
Finally, the edges of the corresponding shaded regions are defined by the values measured on either upstream regions, e.g., the upper extent will be the average reconnection rate determined from the upper half of the current sheet ($\sigma_2$ region) and the lower extent of the shaded region will be the averaged value below the current sheet ($\sigma_1$ region). Considering that $v_1\sim v_2$ for small values of $k_BT/mc^2 \ll \sigma_1,\ \sigma_2$, we expect the shaded regions to have a limited extent especially as $\sigma_2 \rightarrow \sigma_1$.
The simulation measurements were taken at $400\omega_{p}^{-1}$. This is not the longest timescale for all runs, but we choose this time as a representative steady-state snapshot, well after reconnection has started, but before island size inhibits efficient reconnection.

Symmetric simulations in Fig.~\ref{combined} yield results consistent with predictions from previous studies; the reconnection rate increases from the standard non-relativistic 0.1 value, up to almost nearly 1 for simulations with $\sigma \gg 1$ \cite{takahashi+11,liu+15}.
For the asymmetric simulations, $\sigma_1$ is fixed to 10 and $\sigma_2$ is varied over $\sim 4$ orders of magnitude.
For the asymmetric simulations, the reconnection rates are set by the weaker-$\sigma$ regions. 
For $\sigma_2 < \sigma_1$, asymmetric reconnection rates are comparable to the symmetric $\sigma_2$-counterparts, and for $\sigma_2 > \sigma_1$, the rate becomes insensitive to increases in $\sigma_2$.
This can primarily be attributed to the exhaust field lines becoming mass loaded by the weaker $\sigma_1$ (i.e., larger density) inflow. 
From these considerations, our simulations verify that the reconnection rate is quenched by the side of the exhaust with the weaker magnetization parameter, as predicted by Eq.~\ref{quench}, even in the extreme limit where $n^{\prime}_2 \rightarrow 0$ or vacuum, i.e., $\sigma_2 \rightarrow \infty$ (as shown in the Supplemental Material). 

\paragraph{\underline{Nonthermal Particle Spectra}}

\begin{figure}
\centering
\includegraphics[width=0.5\textwidth,clip=false, trim= 0 0 0 0]{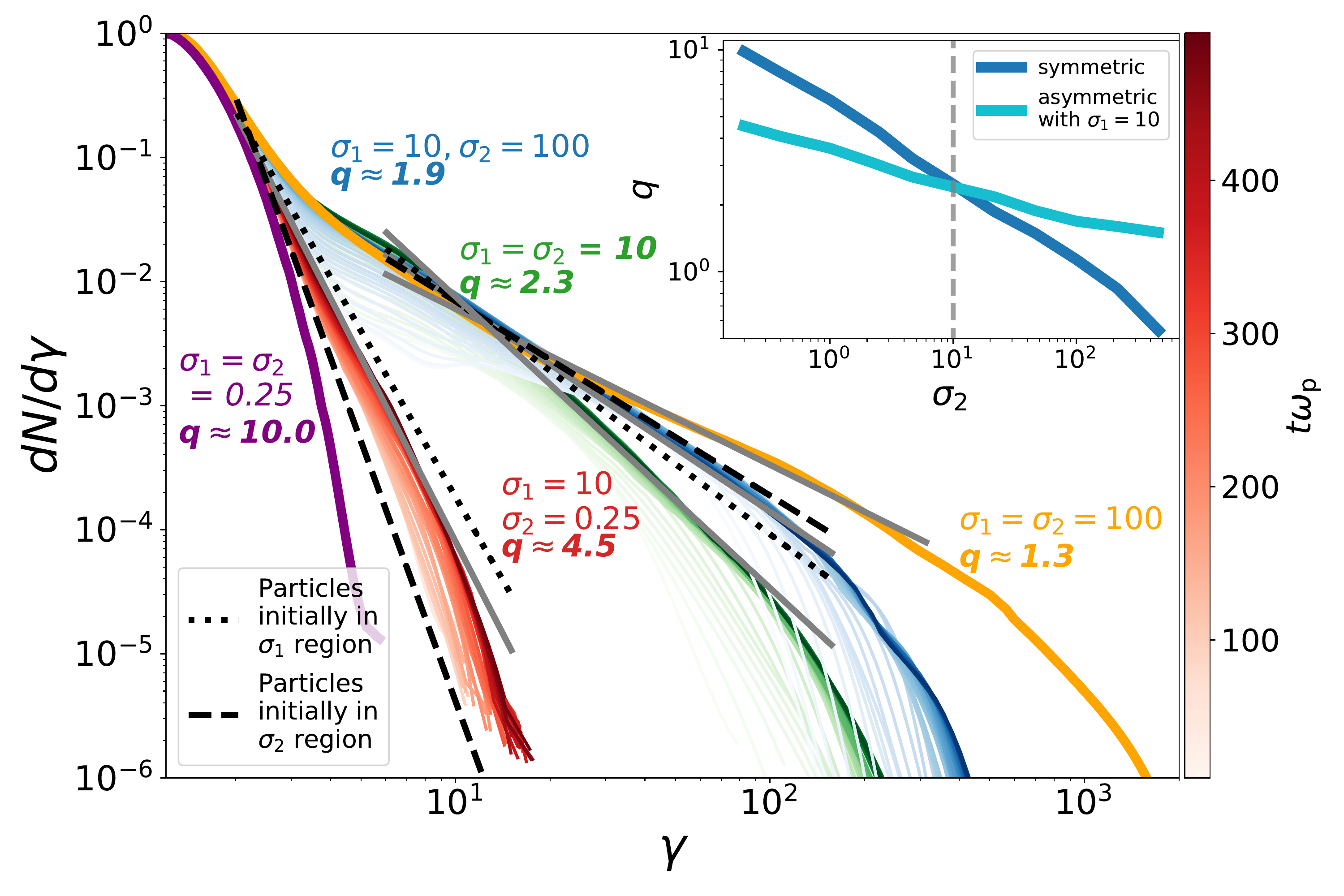}
\caption{Time evolution of particle energy spectra for symmetric and asymmetric reconnection with different magnetization parameters. The time stamp of each distribution is color coded based on the color bar shade for one representative example ($\sigma_1 = 10, \sigma_2 = 0.25$).
The spectrum tends towards a power law slope as time progresses. The dashed and dotted lines show the best fit values for slopes of the power law spectra based on particles originating from the lower and upper inflow regions respectively, while the gray solid lines show the best fit for the combined population. The upper right plot shows the dependence of the slope for different magnetization parameters for both symmetric and asymmetric reconnection.
}
\label{asym}
\end{figure}

To examine particle acceleration, we show the energy distribution functions in Fig.~\ref{asym} for different symmetric and asymmetric conditions at different times, averaged over the entire domain, with the time stamp of representative distributions, color coded based on the color bar shade. We show the distributions for five different characteristic example simulations in Fig.~\ref{asym} that exhibit nonthermal, extended tails reaching ultra-relativistic energies.
The spectra with the red, blue, and green color correspond to $\sigma_2 = 0.25,\ 10,\ \&\ 100$ respectively, with $\sigma_1$ fixed to 10 and are fitted at $t\omega_p \sim 450$ with a spectral slope $q \equiv -d \rm \log{N}/\log{\gamma}$.
We also show the spectra for the symmetric cases $\sigma = 0.25$ and $\sigma = 100$ (the purple and orange lines, respectively) at $t\omega_p \sim 450$ with their power law fit.

Asymmetric simulations develop power-law spectral slopes between those of their symmetric counterparts.
Following similar trends, we find that the spectra associated with $\sigma_2 = 0.25$ (red curves) are much steeper than those with a symmetric $\sigma = 10$ prescription, and can conclude that a power law distribution essentially does not form. The symmetric $\sigma = 0.25$ case has the steepest distribution as expected.

The inset of Fig.~\ref{asym} shows the power law slopes for symmetric and asymmetric ($\sigma_1=10$ for asymmetric cases) simulations over a range of $\sigma$'s and exhibits the same trend for a range of different magnetizations. 
Overall, it appears from these simulations that for asymmetric reconnection, the accelerated particle spectra are steepened relative to the larger $\sigma$ side, and hardened relative to the weaker $\sigma$ side.

Finally, for spectra associated with asymmetric reconnection, we separately present the spectra of positrons coming from either the upper or lower half of the domain; dotted lines correspond to particles initialized on the lower half of the current sheet ($\sigma_1$), and dashed lines for particles from the upper domain ($\sigma_2$). We note that particles starting in the higher $\sigma$ regions have a flatter slope and tend towards higher energies as can be seen in more detail in Fig.~\ref{asym}.

\paragraph{\underline{Conclusions}}
In this work we present the first examination
of relativistic, magnetic reconnection, with asymmetric inflowing conditions.
We consider systems with asymmetric densities and magnetic fields but with a constant, non-relativistic temperature for simplicity.
Using conservation of mass and energy, we derive a scaling prediction for the outgoing bulk flow Lorentz factor, and show that the prediction is consistent with standard limiting cases.
For systems with a large asymmetry in  magnetization, the evolution of reconnection is determined by the smaller $\sigma$ value (Fig.~\ref{combined}). 

The predictions are tested using a survey of 2.5D fully kinetic particle-in-cell simulations performed with TRISTAN-MP \citep{bunemann93,spitkovsky05}.
The scaling relations are found to accurately predict the outflow/reconnection rate of simulations.
We anecdotally find that asymmetric simulations generate less plasmoids than their symmetric counterparts.

We also consider the production of nonthermal particles in the simulations, finding that extended power-law distributions can be produced by asymmetric reconnection, with a
slope that depends on the magnetization of both inflowing plasmas.
Each asymmetric simulation develops a power-law distribution where the spectral slope falls between the slopes of the two symmetric counterparts.

While we anticipate these results to extend to fully 3D systems---as has been shown for the scaling in non-relativistic simulations \cite{Daughton+14,Li+20} and the acceleration in relativistic simulations \cite{sironi+sp14,guo+14}---it should be noted that reconnection in 3D is susceptible to other instabilities that can potentially affect the scaling properties and particle acceleration \cite{Price+16,zhang+21,zhanghao+21}.
This work makes the first step towards understanding asymmetric relativistic magnetic reconnection, but future work is needed to characterize the full 3D nature of the problem.

\section*{Acknowledgements}
Simulations were performed on computational resources provided by the University of Chicago Research Computing Center and XSEDE TACC (TG-AST180008). 
C.C.H acknowledges support from NSF FDSS grant AGS-1936393,
L.S. acknowledges support from the Cottrell Scholars Award, NASA 80NSSC20K1556, NSF PHY-1903412, DoE DE-SC0021254 and NSF AST-2108201,
D.C. acknowledges support from NASA grant 80NSSC18K1218 and NSF grant PHY-2010240,
and M.S. acknowledges NSF grant AGS-2024198
NASA grant 80NSSC20K0198.

\bibliographystyle{apsrmp}
\bibliography{RAreconnection}

\clearpage

\section*{Supplementary Material}

\subsection*{Reconnection Region}
The rate at which magnetic energy is converted during magnetic reconnection can be estimated by considering the diffusion region around a given X-line. The Figure below shows the expected layout of the X-line with the different physical parameters defined in the text.

\begin{figure}[h!]
\centering
\includegraphics[width=0.45\textwidth,clip=True, trim= 0 90 0 100]{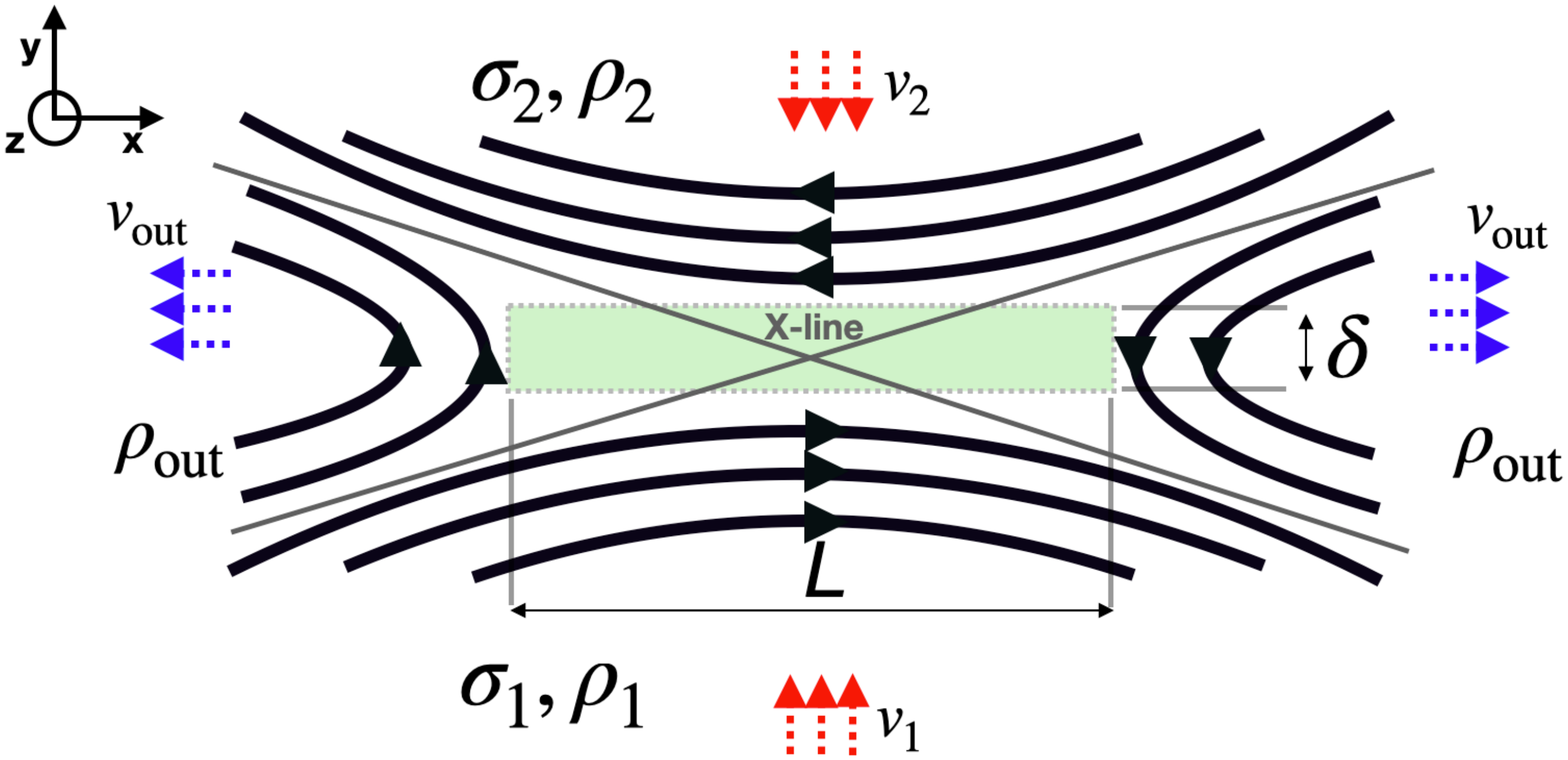}
\captionsetup{labelformat=empty}
\caption{Reconnection region with variables defined in the text.}
\end{figure}

\subsection*{Simulation Setup}
To test the scaling predictions outlined in the text, we perform 2.5D (2D real space, 3D velocity space) particle-in-cell (PIC) simulations of electron-positron (pair-plasma) reconnection with TRISTAN-MP \cite{bunemann93,spitkovsky05}.
The simulations are performed in the $x,\ y$ plane with the reconnecting magnetic field pointing in the $\pm \hat{x}$ direction.
The simulations are periodic in the outflow direction and open in the inflow direction with receding injectors to allow the simulations to continuously feed in upstream plasma and magnetic flux; this technique has been used in various other studies of relativistic reconnection to efficiently reduce the simulation's computational cost \cite{sironi+sp14,sironi+16,rowan+17,ball+18}.
The magnetic field is initialized with ${\bf B}^\prime = [0.5(B^\prime_2 + B^\prime_1) \tanh{(2\pi y/\delta)} + 0.5(B^\prime_2-B^\prime_1)]{\bf \hat{x}}$;
the field on both sides of the current sheet is characterized by the magnetization $\sigma = B^{\prime 2}/4\pi m n^\prime c^2$.
The plasma is comprised of electron-positron pairs, typical of high-energy astrophysical environments such as pulsar winds and AGN jets \cite[e.g.][]{sturrock71,wardle+98}.
Throughout the simulations, no notable differences between electron and positron properties are observed.
The simulation dimensions are $840~c/\omega_p$ in the periodic direction ($x$),
where $\omega_p$ is the electron plasma frequency based on the density of the lower half of the simulation, $\omega_p = \sqrt{4\pi n^\prime_1 e^2/m}$.
We check for convergence by performing additional test simulations with $x$ domains up to $3360~c/\omega_p$ (An example with $1200~c/\omega_p$ is shown in Figure~1 of the text), with no significant differences found.
 We also run the simulations with better particle statistics by increasing the number of particles per cell by an order of magnitude and noted no difference in the scaling and the non-thermal spectra.
The reconnection process is initialized by creating a localized low temperature perturbation in the center of the simulation, thereby collapsing the current sheet in a predetermined location and shortening the time it takes for a dominant X-line to form, as described in \cite{sironi+16}.

\subsection*{Simulation Normalizations}
Unless otherwise stated in the text the simulations are initialized as follows: The length of the simulation in the x direction is $420~c/\omega_p$, based on the density of the subscript 1 side (lower half of the simulation).
Each simulation uses four positrons+electrons per cell with 20 cells per $c/\omega_{p}$. The magnetization parameter is defined as
$\sigma = B^{2}/4\pi \rho c^2$.
The time step is set by the speed of light, $0.45 \Delta x/\Delta t = c$.
The electrons and positrons are initialized with a constant temperature, $\Delta \gamma \equiv k_BT^\prime/mc^2 = 0.25 $.
The parameters that change between simulations are the lower magnetization ($\sigma_1$), the upper magnetization ($\sigma_2$), the upper to lower mass density ratio ($\rho_2/\rho_1$) and the upper to lower magnetic field strength ratio ($B_2/B_1$).
These values are given in the table below.
\begin{table}[h]
    \centering
\begin{tabular}{|c|c|c|c|}
\hline \rowcolor{LightCyan}
$\sigma_1$ &
$\sigma_2$ &
$\rho_2/\rho_1$ &
$B_2/B_1$  \\ \hline
0.22 & 0.22 & 1 & 1      \\ \hline
0.46 & 0.46 & 1 & 1      \\ \hline
1 & 1 & 1 & 1            \\ \hline
2.15 & 2.15 & 1 & 1      \\ \hline
4.64 & 4.64 & 1 & 1      \\ \hline
10 & 10 & 1 & 1          \\ \hline
21.5 & 21.5 & 1 & 1      \\ \hline
46.41 & 46.41 & 1 & 1    \\ \hline
100 & 100 & 1 & 1        \\ \hline
215 & 215 & 1 & 1        \\ \hline
464 & 464 & 1 & 1        \\ \hline
10 & 0.1 & 10 & 0.316    \\ \hline
10 & 0.22 & 9.01 & 0.445 \\ \hline
10 & 0.25 & 8.8 & 0.469  \\ \hline
10 & 0.46 & 7.53 & 0.589 \\ \hline
10 & 1 & 5.5 & 0.742     \\ \hline
10 & 2.2 & 3.437 & 0.869 \\ \hline
10 & 4.6 & 1.964 & 0.95  \\ \hline
10 & 22 & 0.478 & 1.025  \\ \hline
10 & 46 & 0.234 & 1.038  \\ \hline
10 & 100 & 0.109 & 1.043 \\ \hline
10 & 220 & 0.0497 & 1.046\\ \hline
10 & 460 & 0.0238 & 1.046\\ \hline

\end{tabular}
\end{table}


\subsection*{Simulations with $\sigma_2 \rightarrow \infty$}

\begin{figure}[h!]
\centering
\includegraphics[width=0.5\textwidth,clip=True, trim= 0 0 0 0]{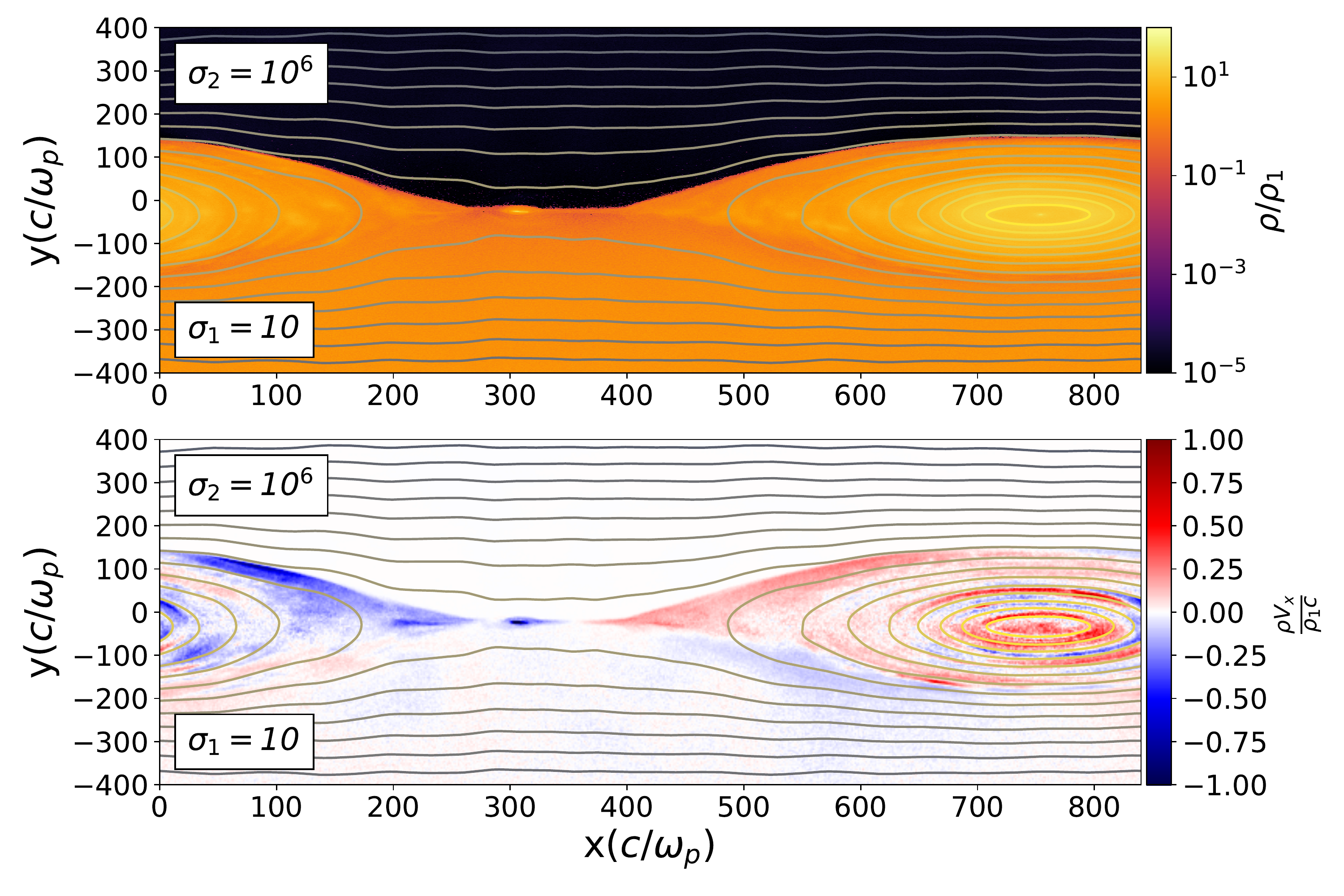}
\captionsetup{labelformat=empty}
\caption{Same as Fig. 1 in the main text but for $\sigma_1 = 10$ and $\sigma_2 \rightarrow \infty$. We can see that reconnection occurs even if one side is in vacuum.}
\end{figure}

We present below a unique extension of this work that would be inadequately described with previous theories of relativistic or asymmetric reconnection. We consider an environment where in one of the reconnecting regions $\sigma \rightarrow \infty$ (de facto vacuum). If one were to analyze this system employing the existing theories of magnetic reconnection, one would face two significant obstacles. First, in the relativistic symmetric framework, the lack of plasma implies that no magnetic energy could be dissipated, resulting in insignificant particle energization and no reconnection for that matter. Second, in the standard asymmetric framework, $V_A \rightarrow \infty$, where $V_A$ is the Alfvén speed on the vacuum side, which is an inherently inaccurate description of the system. However, the model presented in this manuscript is applicable to this extreme regime and predicts that reconnection will occur with a given outflow and reconnection rate. 
We also recover a power law spectrum in this simulation, and observe the nonthermal particle spectrum to be enhanced by reconnecting with vacuum.

\begin{figure}[h!]
\centering
\includegraphics[width=0.5\textwidth,clip=True, trim= 0 0 0 0]{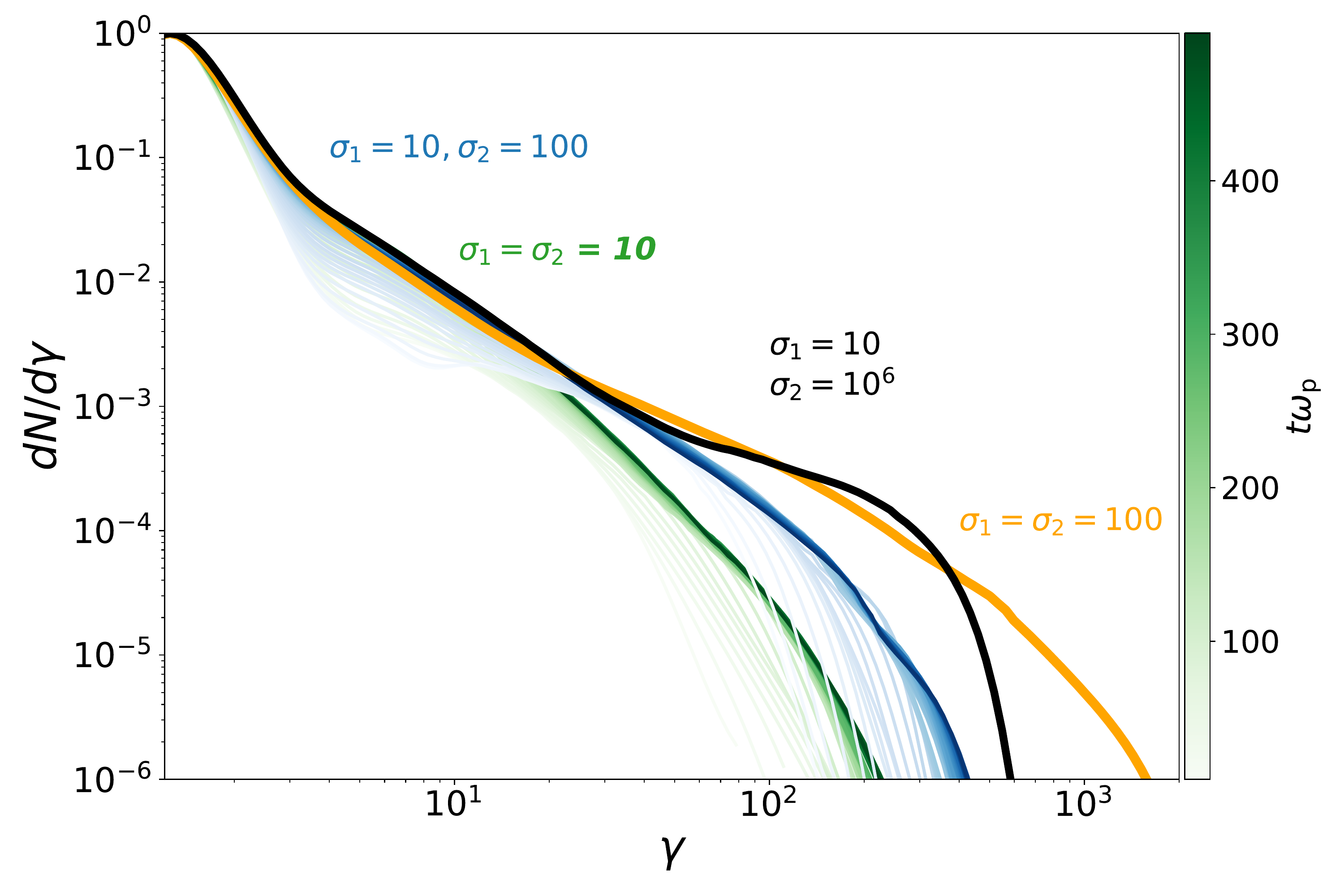}
\captionsetup{labelformat=empty}
\caption{Same as Fig.3 in the main text but including the particle energy spectrum for a system with $\sigma_1 = 10$ and $\sigma_2 \rightarrow \infty$.}
\end{figure}

\end{document}